\documentclass[twocolumn,amsmath,amssymb,prb,floatfix]{revtex4}
\usepackage{psfig}
\makeatletter
\usepackage{graphicx}% Include figure files
\begin{document}

\title{Exchange interactions in (ZnMn)Se}
\author{
L.M. Sandratskii\thanks{lsandr@mpi-halle.de}}

\affiliation{Max-Planck Institut f\"ur Mikrostrukturphysik, D-06120 Halle, Germany }
\begin{abstract}
One of the remarkable properties of the II-VI diluted magnetic semiconductor (ZnMn)Se 
is the giant spin splitting of the valence band states under application of the 
magnetic field (giant Zeeman splitting). 
This splitting reveals strong exchange interaction between Mn moments and
semiconductor states. 
On the other hand, no magnetic phase transition has been observed 
for systems with small Mn content up to very low temperatures. The latter property shows
weakness of the exchange interaction between Mn moments.
In this paper, the local density approximation (LDA) and the LDA+U techniques are 
employed to study exchange interactions in (ZnMn)Se. 
Supercell and frozen-magnon approaches applied earlier 
to III-V diluted magnetic semiconductors 
are used. It is found that both LDA and LDA+U 
describe successfully the combination of the strong Zeeman splitting 
and weak interatomic exchange. However, the physical pictures provided by
two techniques differ strongly.
A detailed analysis shows that the LDA+U method provides the description
of the system which is much closer to the experimental data.
\end{abstract}
\maketitle

\section{Introduction}

The perspective of using the spin of electrons in the 
semiconductor devices promises to revolutionize modern electronics. \cite{ohno_science}
A necessary
component of a spintronic process is the spin-injection into semiconductor.
This demand created a need for ferromagnetic materials on the semiconductor basis
with strong spin-polarization of the carriers and high Curie temperature. 
After recent discovery \cite{ohno_gaas} of ferromagnetism in (GaMn)As with Curie 
temperature as high as 110 K much attention is devoted to the study of the 
diluted III-V semiconductors as possible sources of the 
spin-polarized electrons. Also the interest
to more traditional II-VI semiconductors has been revived since, first, 
the theoretical studies shaw that
the II-VI systems possess the potential 
for high Curie temperature \cite{diohma_science,sato_rev}, second, the study of the II-VI 
diluted magnetic semiconductors (DMS)  
deepens understanding of the exchange interactions in other types of 
DMS \cite{kacman_rev,dietl_rev,sato_rev} and, third, the
II-VI systems, in particular (ZnMn)Se, are used in spin-injection experiments
as a source of spin-polarized
charge carriers. \cite{goscri,jopabe,oehuha}

In (ZnMn)Se with low Mn concentration
no magnetic ordering has been experimentally 
detected up to very low temperatures. \cite{furdyna}
However, the application of magnetic field
leads to the observation of the so-called giant Zeeman splitting for the states of
the semiconductor matrix. Because of this large spin splitting 
(ZnMn)Se is an efficient source of highly polarized carriers. \cite{goscri,jopabe}

A commonly accepted explanation of the giant Zeeman splitting in (ZnMn)Se relies
on the following picture. In the system there is a strong 
exchange interaction between the Mn 3d states and the  
states of the semiconductor. This interaction does not, however,
lead to the ordering of the Mn moments.
The application of a magnetic field aligns 
the Mn moments and results in the observation of a giant spin-splitting. 
The term 'giant' arises here from the comparison of two energy scales.
The characteristic energy of the magnetostatic interaction of the magnetic field of 
1 T with the spin of 1$\mu_B$ amounts to 0.004 mRy and is up to five orders of 
magnitude smaller than the splittings detected experimentally. \cite{goscri,furdyna}
The coexistence of a very strong exchange between the Mn 3d and semiconductor states
and a very low temperature of the magnetic phase transition makes the II-VI 
DMS an interesting laboratory for studying the physics of exchange interactions. 

Much efforts has been paid to the theoretical studies of the 
DMS of the II-VI type (see, e.g., reviews \cite{furdyna,kacman_rev,dietl_rev,sato_rev}). 
Most of these studies are based on a model-Hamiltonian approach. 
[See, e.g., Ref. \cite{dubhwo} for the model of bound magnetic polarons
and Refs. \cite{dietl_rev,feciwa} for the Zener model. The latter can be considered
as a continuous-medium limit of the well-known Rudermann-Kittel-Kasuya-Yosida (RKKY) 
approach. Spalek et al \cite{spalek_86} used Anderson's approach to superexchange
and have shown that few adjustable parameters 
of the theory are sufficient to describe the interplay between different types
of exchange interactions in the system. Reach experimental information 
(e.g. \cite{furdyna_88} in the case of (ZnMn)Se) 
is helpful in the selection of the values of the parameters.] 

The developments in the methods of the density functional theory (DFT) 
accompanied by fast increasing computer power
allow now for parameter-free calculation of the electronic properties of very
complex systems. 
In the case of II-VI DMS the number of the DFT 
studies of the exchange interactions
is still very restricted (for exceptions see, e.g., Refs. \cite{lahaeh,wezu,sato_rev}). 
Most of the calculations have been
performed with the use of a virtual crystal approximation or a single-site
coherent-potential approximation (CPA).\cite{lahaeh,sato_rev}  These schemes are convenient 
and efficient in the
investigation of the systems with varying concentration of impurities. They, however, do not
take into account the atomic short-range order and the increasing distance between 
impurities with decreasing impurity concentration. Therefore it is important to combine 
the calculations within the virtual-crystal and CPA techniques with the studies 
making more detailed account for the positions of the impurity atoms. Such 
studies can be performed with the use of large supercells of the semiconductor crystals. 
\cite{zhao,sanvito_rev,sabr02_1,sabr_condmat} 
The aim of this paper is the investigation of the exchange interactions in 
(ZnMn)Se on the basis of the supercell approach. 

An important question concerns the role of the intraatomic correlations 
in the Mn 3d states. Within the model-Hamiltonian approaches the 3d states are
usually considered as atomic-like and strongly correlated. This treatment is 
rather different from the treatment within LDA.
To study the role of the electron correlations in the Mn3d shell we use the
LDA+U method \cite{LDA+U_93} designed to take into account the on-site
Coulomb interaction $U$.  

The purpose of this paper is two-fold. On the one hand, by detailed DFT calculation of the 
electronic properties of (ZnMn)Se we aim to provide deeper insight into the
physics of the system. On the other hand, by comparison with experimental data of the
calculational results we aim to draw the conclusion which of the two approaches,
LDA or LDA+U, provides better description of (ZnMn)Se. Both components of the purpose are 
of strong importance for future studies of DMS within the DFT.
 
\section{Calculational approach}

In the calculations we follow the scheme described in  Ref. \cite{sabr02_1}.
This scheme is based on the supersell approach  where one of the 
Zn atoms in a supercell of zinc-blende ZnSe is replaced by the Mn atom.
The calculations are performed for four values of the concentration $x$:
0.25, 0.125, 0.0625, and 0.03125.

The calculations were carried out with the augmented spherical waves \cite{wikuge} (ASW) 
method within the LDA and LDA+U approaches. In all calculations the lattice parameter 
was chosen to be equal to the experimental lattice parameter of ZnSe.
Two empty spheres per formula unit have been used in the calculations.
The positions of empty spheres are (0.5, 0.5, 0.5) and (0.75, 0.75, 0.75).
Radii of all atomic spheres were chosen to be equal. 
Depending on the concentration of Mn, the super cell is cubic
(x=25\%, 
$a\times a\times a$, and x=3.125\%, $2a\times 2a\times 2a$) 
or tetragonal 
(x=12.5\%, 
$a\times a \times 2a$ and 6.25\%, $2a\times 2a\times a$).

The LDA+U calculations were performed with $U=0.3Ry$. 
\cite{U_comment} 

\subsection{Spin-projected densities of states}

\begin{figure}
\caption{
Two scenarios of the relation between the net magnetization due to the
localized moments and the exchange splitting of the valence state. In the
mean-field (Stoner-like) scenario the splitting is proportional to the net
magnetization. The distance between 
wave lines shows schematically the value of the exchange splitting.
In the non-mean-field (non-Stoner) scenario the spin of the valence electrons follows
locally the direction of the Mn moments. In this case the exchange splitting 
is present also in the case of zero net magnetization.
 \label{fig_stoner}}
\centerline{\includegraphics[width=8cm,angle=-90]{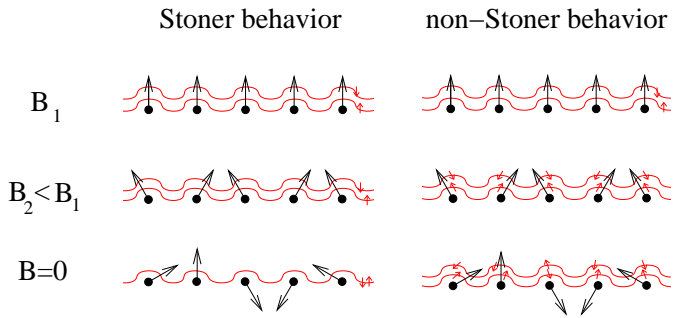}}
\end{figure}

The spin splitting of the valence-band states is a result of the
interaction between these states and the Mn3d states.
The directions of the Mn moments are disordered in the absence of magnetic field and
become increasingly ordered with increasing value of the applied field.
There are two possible scenarios for the relation between the ordering of the Mn 
moments and the spin splitting of the valence band states (Fig. \ref{fig_stoner}). First scenario
is a mean-field (Stoner-like) type of the relation.
In this case the valence-band states experience an average exchange field of the
Mn moments that is proportional to the net magnetization in the Mn subsystem.
Complete disordering of the Mn moments leads to the disappearance of the net magnetization
and, respectively, of the spin splitting. 
In the second scenario, the spin of the valence states follows 
locally the spins of the Mn atoms. Therefore, the spin polarization and exchange splitting 
do not disappear with disappearance of the net magnetization. 
An experimental example of the non-Stoner behavior of the exchange splitting is 
discussed by Kisker \cite{kisker} 
for the case of iron with thermally disordered atomic moments.
In (ZnMn)Se the experimental data is treated in favor of the mean-field
scenario. \cite{furdyna}

To study both scenarios within the DFT
we will need to calculate densities of states (DOS)
projected on different spin-quantization axes.
The calculation is performed as follows.
The wave function of a given electron state is considered to be two-component spinor
$\left(\begin{array}{c}\psi_1 ({\bf r}) \\ \psi_2 ({\bf r})\end{array}\right)$.
The spin components $\psi_1 ({\bf r})$ and  $\psi_2 ({\bf r})$ 
are written with respect to a chosen axis. 
The spin-quantization axis does not change within atomic spheres
but can vary from atom to atom. The integral
\begin{equation}
\label{eq_n_at_s}
n^s_{i{\bf k}\nu} =\int_{\Omega_\nu} d{\bf r} |\psi_{i{\bf k}s} ({\bf r})|^2
\end{equation} 
gives the part of the state 
$\psi_{i{\bf k}}$ corresponding to atom $\nu$ and  spin-projection $s$. 
Here {\bf k} is the wave vector, $i$ numbers the energy bands.
The integration is carried out over the $\nu$th atomic sphere.
The wave functions are normalized in the unit cell:
$\sum_{s\nu}n^s_{i{\bf k}\nu} =1$.
The partial DOS for given $\nu$ and $s$ is given by formula
\begin{equation}
\label{eq_N_at_s}
N^s_\nu(\varepsilon)=\frac{1}{\Omega_{BZ}}\sum_i
\int_{BZ}d{\bf k}\: n^s_{i{\bf k}\nu}\:\delta(\varepsilon-\varepsilon_{i{\bf k}})
\end{equation} 
where $\Omega_{BZ}$ is the volume of the Brillouin zone (BZ).
To calculate the partial DOS with respect to another
quantization axis the electron wave functions are subjected to the 
transformation \cite{adv_phys} 
\begin{equation}
\label{eq_U_trans}
\left(\begin{array}{c}\psi_1^\prime ({\bf r}) \\ \psi_2^\prime ({\bf r})\end{array}\right)
\:=\:U^\dag_\nu
\:\left(\begin{array}{c}\psi_1 ({\bf r}) \\ \psi_2 ({\bf r})\end{array}\right)
,\:\:\:\:{\bf r}\in \Omega_\nu
\end{equation} 
where $U_\nu$ is the spin-$\frac{1}{2}$ transformation matrix corresponding to the
rotation of the axis of the $\nu$th atom. 
Next, the procedure defined by Eqs. (\ref{eq_n_at_s}-\ref{eq_N_at_s})
is performed for the spinor components of the transformed functions (\ref{eq_U_trans}).

\subsection{Interatomic exchange parameters and Curie temperature}

To describe the exchange interactions between Mn moments
we use an effective Heisenberg Hamiltonian of classical spins
\begin{equation}
\label{eq:hamiltonian}
H_{eff}=-\sum_{i\ne j} J_{ij} {\bf e}_i\cdot {\bf e}_j
\end{equation}
where $J_{ij}$ is an exchange interaction between two Mn sites $(i,j)$
and ${\bf e}_i$ is the unit vector pointing in the direction 
of the magnetic moment at site $i$. 

To estimate the parameters of the Mn-Mn exchange interaction we
perform calculation for the following frozen-magnon 
\cite{haespe,nikl}  configurations:
\begin{equation}
\theta_i=const, \:\: \phi_i={\bf q \cdot R}_i
\end{equation}
where $\theta_i$ and $\phi_i$ are the polar and azimuthal angles of vector
${\bf e}_i$, ${\bf R}_i$ is the position of the $i$th Mn atom. 
The directions of the induced moments in the atomic spheres of Ga and As
and in the empty spheres were kept to be parallel to the $z$ axis. 

It can be shown that within the Heisenberg model~(\ref{eq:hamiltonian})
the energy of such configurations can be represented in the form
\begin{equation}
\label{eq:e_of_q}
E(\theta,{\bf q})=E_0(\theta)-\theta^2 J({\bf q})
\end{equation}
where $E_0$ does not depend on {\bf q} and $J({\bf q})$ is the 
Fourier transform of the parameters of the exchange interaction between 
pairs of Mn atoms:
\begin{equation}
\label{eq:J_q}
J({\bf q})=\sum_{j\ne0} J_{0j}\:\exp(i{\bf q\cdot R_{0j}}).
\end{equation}
In Eq. (\ref{eq:e_of_q}) angle $\theta$ is assumed to be small.
Using $J({\bf q})$ one can estimate the energies of the spin-wave 
excitations \cite{comm_proport}:
\begin{equation}
\label{eq_ome}
\omega({\bf q})=\frac{4}{M} [J({\bf 0})-J({\bf q})]=
\frac{4}{M} \frac{E(\theta,{\bf q})-E(\theta,{\bf 0})}{\theta^2}
\end{equation} 
where $M$ is the atomic moment of the Mn atom.
Performing back Fourier transformation we obtain the parameters of 
the exchange interaction between Mn atoms:
\begin{equation}
\label{eq:J_0j}
J_{0j}=\frac{1}{N}\sum_{\bf q} \exp(-i{\bf q\cdot R_{0j}})J({\bf q}).
\end{equation}

The Curie temperature was estimated in the mean-field (MF) approximation
\begin{equation}
\label{eq:Tc_MF}
k_BT_C^{MF}=\frac{2}{3}\sum_{j\ne0}J_{0j}
\end{equation}

We use rigid band 
approach to calculate the exchange parameters and Curie temperature
for different electron occupations. 
We assume that the electron structure calculated for a DMS with a given
concentration of the 3d impurity is basically preserved in the
presence of defects. The main difference
is in the occupation of the bands and, respectively, 
in the position of the Fermi level.

\section{Calculational results}
\subsection{Densities of states and exchange splittings}
\label{sec_split}
\subsubsection{LDA}

\begin{figure}
\caption{The DOS of Zn$_{1-x}$Mn$_{x}$Se. The DOS is given per 
unit cell of the zinc-blende crystal structure. The DOS above(below)
the abscissas axis corresponds to the spin-up(down) states. 
 \label{fig_DOS}}
\includegraphics[width=8cm]{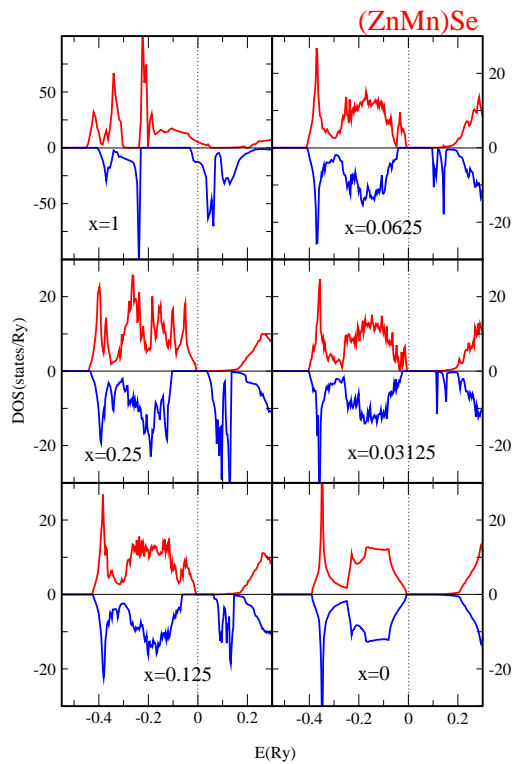}
\end{figure}

We begin the discussion of 
the calculational results with consideration of 
the DOS of the ferromagnetic (ZnMn)Se
(Fig. \ref{fig_DOS}). 
Compared with pure ZnSe, the replacement of a Zn atom by a Mn atom adds
five 3d spin-up energy bands
to the valence band of the system. Since the
number of Mn 3d electrons is also five no carriers appear either
in the valence band or in the conduction band. 

\begin{table}
\caption{Magnetic moments in Zn$_{1-x}$Mn$_{x}$Se. There are shown the
Mn moment, the induced moment on the nearest Se atoms, and the 
magnetic moment of the super cell.
All moments are in units of $\mu_B$. 
\label{tab:moments}}
\begin{tabular}{lcccc}
&\multicolumn{4}{c}{x}\\
&0.25&0.125&0.0625&0.03125\\
\colrule
&\multicolumn{4}{c}{LDA}\\
Mn&  4.39&   4.39 &       4.39&  4.40 \\
As&  0.07  &  0.06&    0.06 &     0.06 \\
cell&   5.0&        5.00&         5.00&      5.00\\
&\multicolumn{4}{c}{LDA+U}\\
Mn&  4.52&   4.51 &       4.51&  4.51 \\
As&  0.06  &  0.05&    0.05 &     0.05 \\
cell&   5.0&        5.00&         5.00&      5.00\\
\colrule

\end{tabular}
\end{table}
The values of the calculated spin moments 
are collected in Table~\ref{tab:moments}.
The moment in the Mn sphere and the induced moment on the neighboring Se atom 
are practically independent of the Mn concentration. The moment per supercell
is exactly 5$\mu_B$ since there are extra five filled spin-up bands
compared with the nonmagnetic ZnSe.

\begin{figure}
\caption{The partial Mn3d-DOS for Zn$_{1-x}$Mn$_{x}$Se. The DOS is given per 
Mn atom.
\label{fig_dos_Mn}}
\includegraphics[width=8cm]{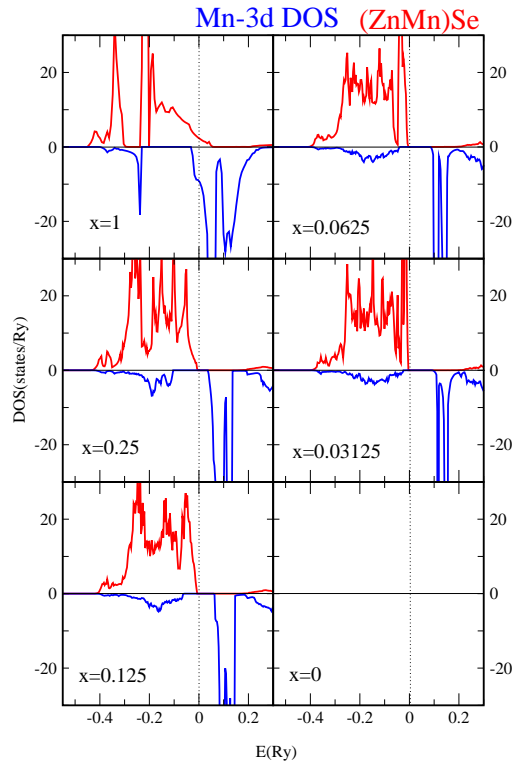}
\end{figure}

At the top of the valence band there is strong negative exchange splitting:
the upper edge of the spin-up DOS
lies higher in energy than the upper edge of the spin-down DOS ( Fig. \ref{fig_DOS}). The spin 
splitting increases with increasing Mn concentration. The origin of this
splitting can be understood from the analysis of the partial Mn3d DOS
(Fig. \ref{fig_dos_Mn}). 
Indeed, a higher energy of the spin-up valence-band edge
results from the contribution of the Mn3d spin-up states. 
On the other hand, there is no strong contribution of the 
Mn spin-down states to the valence band. 
Unoccupied spin-down Mn3d bands lie in the semiconducting gap, close to the 
bottom of the conduction band. 

The exchange splitting of the states at the top 
of the valence band of the ferromagnetic (ZnMn)Se is in qualitative agreement with 
the observation of the giant spin splitting under the application of the
magnetic field. 

\begin{figure}
\caption{Total DOS and partial Mn3d-DOS for the antiferromagnetic Zn$_{1-x}$Mn$_{x}$Se
with $x=3.125\%$. The spin-projected DOS are calculated with respect to the local
atomic quantization axes.
\label{fig_DOS_afm}}
\includegraphics[width=8cm]{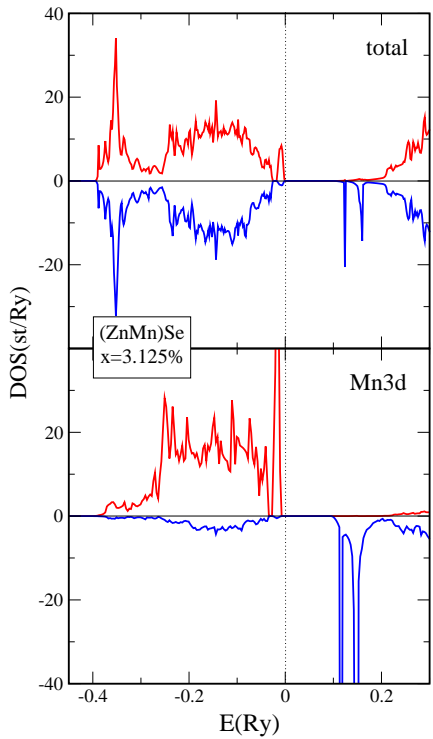}
\end{figure}
To understand the dependence of the electron structure on the magnetic configuration we 
performed calculation for the antiferromagnetic configuration of the Mn moments
for $x=3.125\%$. 
The antiferromagnetic structure is characterized by the largest angle 
between the neighboring Mn moments and, therefore, is 
the state most different from the ferromagnetic one. The comparison 
of the DOS of the ferromagnetic and antiferromagnetic configurations for 
$x=3.125\%$ (Figs. \ref{fig_DOS} and \ref{fig_DOS_afm}) shows that 
they are similar. The main difference in the antiferromagnetic DOS is the 
appearance of a small energy gap 
which separates the upper part of the valence-band states from the rest of the
valence band. 
The 
spin-splitting obtained in the ferromagnetic DOS
is present also in the antiferromagnetic DOS
as the splitting between the top of the separated (impurity) band and the top of 
the valence band. 
Figures \ref{fig_dos_Mn} and
\ref{fig_DOS_afm} show that the partial DOS of the Mn atoms are 
only weakly dependent on the magnetic configuration and preserve fully the
local spin splitting. 
Obviously the main features of the LDA-DOS cannot
be treated in terms of the Stoner-like mean-field picture (Fig. \ref{fig_stoner}). 

To study in more details the relation between  
the electron structure and magnetic configuration we perform 
the analysis of the partial DOS of the atoms of
the semiconductor matrix. As we show below, 
the behavior of the DOS of different atoms ranges from a highly non-Stoner one 
to the behavior well described by the mean-field picture.  

\begin{figure}
\caption{Fragment of the partial DOS of the Se-I atom
at the top of the valence band for Zn$_{0.96875}$Mn$_{0.03125}$Se.
Calculations are performed for seven magnetic configurations with different
net magnetization. The upper part shows the spin-projection on the global $z$ axis.
The lower part gives the spin-projections on the axis parallel to the 
direction of the magnetic moment of the nearest Mn atom. No correlation between 
the spin-splitting and the net magnetization can be established.  
\label{fig_Se_I}}
\centerline{\includegraphics[width=7cm,angle=-90]{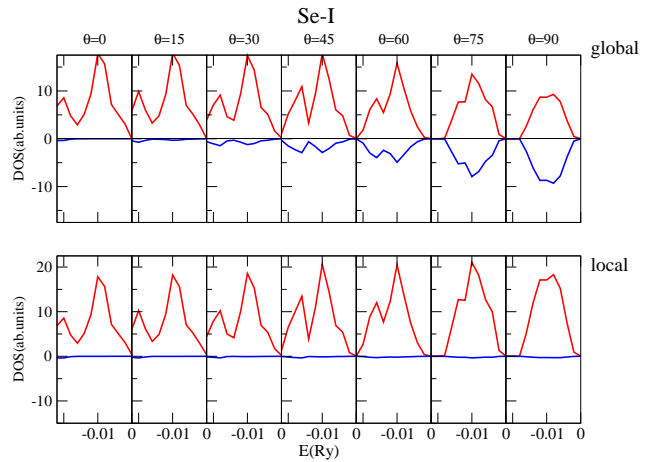}}
\end{figure}
\begin{figure}
\caption{Fragment of the partial DOS of the Se-II atom. See the caption
of Fig. \ref{fig_Se_I} for details.
Short vertical lines in the upper panel show the center of gravity for the 
corresponding spin-projection. 
The decrease of the
spin-splitting with decreasing net magnetization can be established.
\label{fig_Se_II}}
\centerline{\includegraphics[width=7cm,angle=-90]{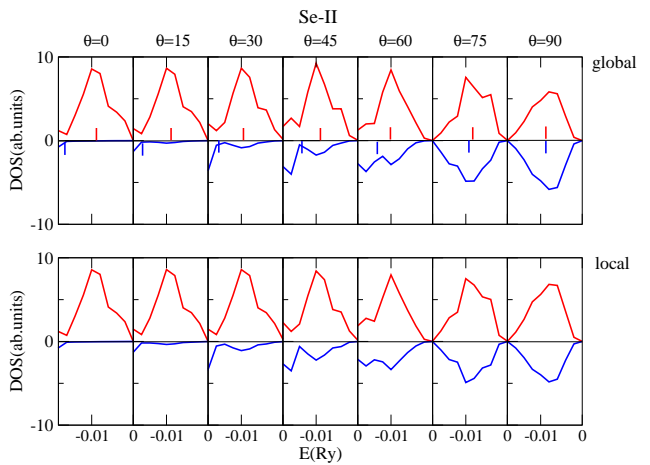}}
\end{figure}
\begin{figure}
\caption{Schematic picture of the calculated magnetic configurations.
Angle $\theta$ assumed the following values: 
0$^\circ$,15$^\circ$,30$^\circ$,45$^\circ$,60$^\circ$,75$^\circ$,90$^\circ$.
Calculations were performed for (ZnMn)Se with the Mn concentration of  
$x=3.125\%$.
\label{fig_noncol_struct}}
\includegraphics[width=8cm,angle=-90]{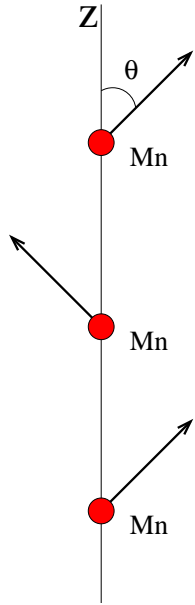}
\end{figure}

In Figs. \ref{fig_Se_I},\ref{fig_Se_II} we present
the partial DOS for two different Se atoms. The DOS is calculated for magnetic
configurations depicted schematically in Fig. \ref{fig_noncol_struct}
where three successive Mn atoms along the $z$ axis are shown.
The angles between neighboring Mn moments vary from
0 to 180$^\circ$ with a step of 30$^\circ$. Correspondingly, the net 
magnetization vary from maximal to zero.

First of the two Se atoms, Se-I, is
at position $(\frac{a}{4},\frac{a}{4}, \frac{a}{4})$
and is the nearest neighbor of the Mn impurity situated at (0,0,0). 
The second atom, Se-II, is at position 
$(\frac{3a}{4},\frac{3a}{4}, \frac{5a}{4})$ and belongs to the 4th coordination
sphere of Se atoms. (To remind, the supercell in this case is a cube of the size $2a$.)  
The spin-projected DOS is presented with respect to two different quantization axes.
The first (local) axis is parallel to the direction of the magnetic moment
of the nearest Mn atom. The second (global) axis is directed along the 
net magnetization. The DOS projected on the local atomic quantization axis
provides, in most cases, better insight into the physics of the system. 
However, in the experiments determining
the spin-splittings the spin-quantization axis is usually the global one.  

The DOS shown in Figs.  \ref{fig_Se_I},\ref{fig_Se_II}
reveal strong difference between two Se atoms.
The spin of the electron states of the Se-I atom follows almost perfectly the 
spin of the neighboring Mn atom (Fig. \ref{fig_Se_I}). 
This is evidenced by the negligibly small local
spin-down DOS. The smallness of the local spin-down DOS holds up to the largest 
angle between Mn moments. 
Considered from the viewpoint of the global quantization axis, both spin-up and spin-down
DOS of Se-I have similar shape but different weights (amplitudes).
The relative weight of the spin-down DOS increases
from zero for $\theta=0^\circ$ to 1 for $\theta=90^\circ$. 
Because of the similarity of the form of the global spin-up and spin-down DOS 
the variation of the DOS with the change of the net magnetization
cannot be treated in terms of the varied spin-splitting.
The changes in the global DOS take the form of the redistribution of the 
weight between the spin-up and spin-down DOS. 
This behavior is principally different 
from the behavior expected within the mean-field picture.

On the other hand, for Se-II we get a  
strong dependence of the local DOS on the magnetic configuration of the Mn moments 
(Fig. \ref{fig_Se_II}).
With increasing angle between Mn moments
the contribution of the spin-down DOS increases. This happens because the spin of the states
of Se-II deviates increasingly from the spin of the nearest Mn atom
responding to the influence of other Mn atoms.
Now, the shape of the spin-up and spin-down DOS calculated with respect to the global
quantization axis differ strongly. To characterize this difference 
in terms of the spin-splitting we calculated the centers of gravity for both
spin-DOS in the energy region at the top of the valence band (Fig. \ref{fig_Se_II}).  
\begin{figure}
\caption{The exchange splitting calculated from the partial DOS of the Se-II atom
as a function of the net magnetization. Negative value of the splitting 
reflects higher energy position of the spin-up states.
The dashed straight line is a guide for the eye.
\label{fig_splitting}}
\centerline{\includegraphics[width=8cm,angle=-90]{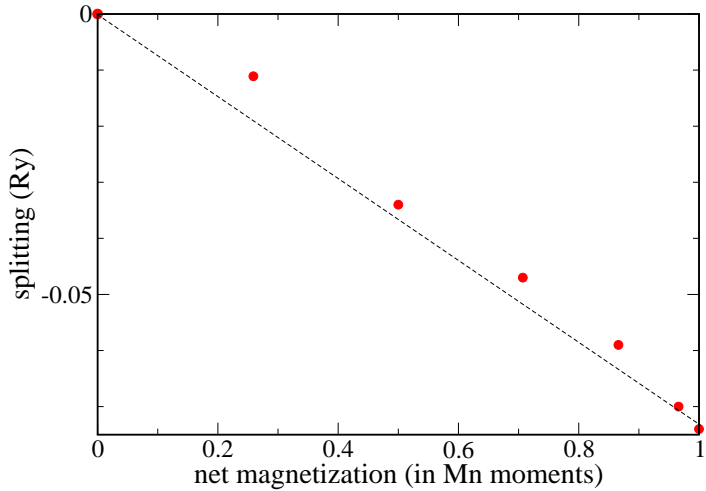}}
\end{figure}
In a good approximation 
the spin-splitting is proportional to the
net magnetization (Fig. \ref{fig_splitting}).
Therefore the properties of the electron structure of Se-II
can be described within the mean-field picture.

Since the relation between the exchange splitting and
net magnetization varies from atom to atom,
in experiments probing different parts of the electron structure this relation 
can appear different.

\begin{figure}
\caption{
The exchange splitting at the top of the valence band ($\Gamma$ point).
Calculations are performed within LDA and LDA+U. The solid line corresponds to the 
experimental value of the $J_{pd}$ parameter. \cite{kacman_rev}
 \label{fig_Jpd}}
\centerline{\includegraphics[width=7cm,angle=-90]{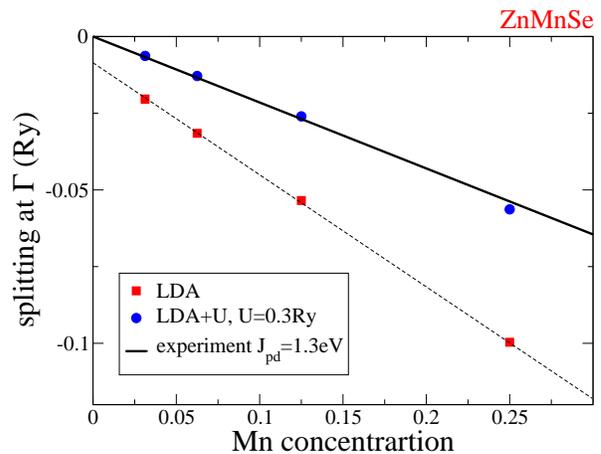}}
\end{figure}
Important role in the experimental determination of the strength of the exchange interaction between the 
Mn moments and valence band states in DMS is played by the magneto-optical  
measurements of the spin splitting at the top of the valence band 
($\Gamma$ point of the Brillouin zone). \cite{furdyna}
In Fig. \ref{fig_Jpd} we show the calculated dependence of the splitting on Mn concentration.
The LDA results for the exchange splitting at the $\Gamma$ point
disagree with experiment in two respects.
First, the mean-field picture which assumes proportionality between the exchange splitting
and the net magnetization in the system does not apply to the LDA results.
The LDA-splitting is well described by a linear function with 
a finite value in the limit of $x \rightarrow 0$.
Such form of the dependence is the result of the presence of the 
Mn3d states at the Fermi level. Second, the exchange splitting is substantially 
larger than the splitting obtained experimentally. Since the LDA results 
cannot be described within the mean-field picture the use of the formula 
\begin{equation}
\label{eq_lpd}
J_{pd}=\frac{\Delta E}{S\:x}
\end{equation}
gives values of the $J_{pd}$ parameter that depend on concentration $x$.
For illustration, the value of $J_{pd}$ obtained for $x=0.125$ according to
Eq. (\ref{eq_lpd}) is about two times larger than the experimental value.   
In Eq. (\ref{eq_lpd}), $\Delta E$ is the exchange splitting and $S$
is the atomic spin moment of Mn.

\begin{figure}
\caption{The DOS of Zn$_{1-x}$Mn$_{x}$Se calculated within the
LDA+U approach with U=0.3Ry. (To compare with Fig. \ref{fig_DOS}
presenting LDA-calculation.)
 \label{fig_DOS_U}}
\includegraphics[width=8cm]{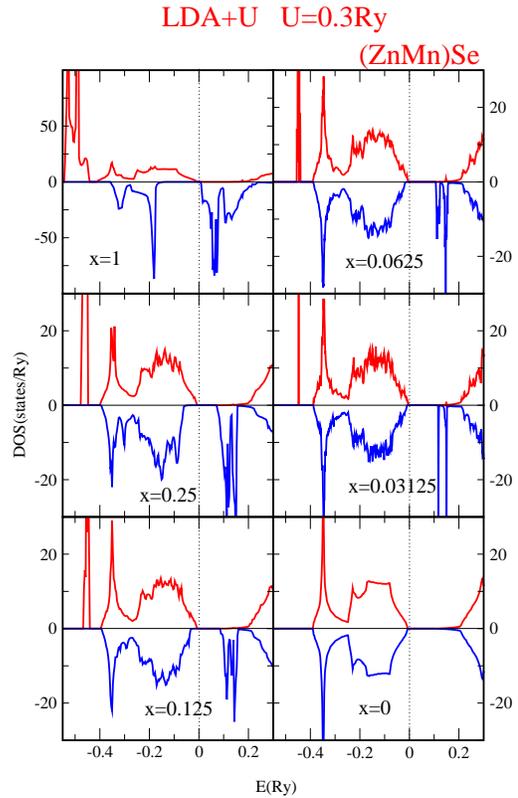}
\end{figure}

\subsubsection{LDA+U}

\begin{figure}
\caption{The partial Mn3d-DOS for Zn$_{1-x}$Mn$_{x}$Se
calculated within the
LDA+U approach with U=0.3Ry. 
(To compare with Fig. \ref{fig_dos_Mn} presenting LDA-calculation.
Notice the change in the scale of the ordinate axis that reflects 
strong change in the DOS.)
\label{fig_dos_U_Mn}}
\includegraphics[width=8cm]{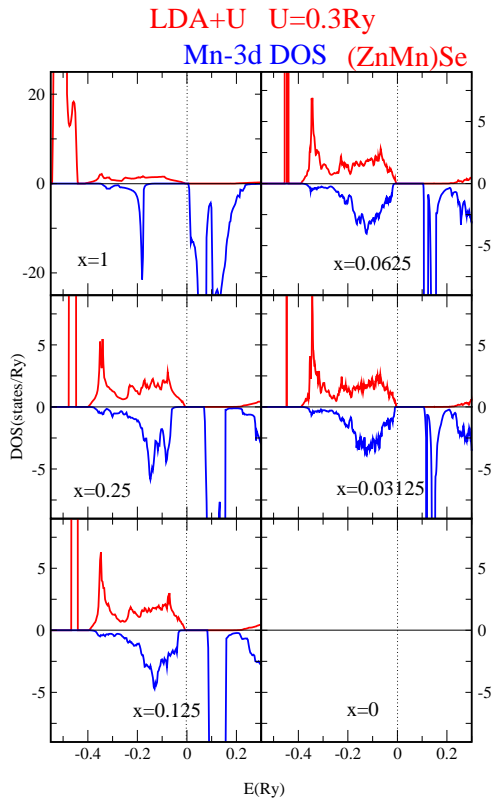}
\end{figure}

Introducing of the Hubbard-U into calculational scheme results
in a strong shift of the Mn3d spin-up states to
lower energies (Fig. \ref{fig_DOS_U},\ref{fig_dos_U_Mn}).
At the top of the valence band there is still an admixture of the Mn3d states.
It is, however, very weak, especially for the spin-up states.
Nevertheless, one can notice the hybridizational repulsion of the valence-band spin-down states
from the spin-down Mn3d states lying in the semiconductor gap. 
This repulsion is an important factor leading, in agreement with experiment, 
to negative exchange splitting at the top of the valence band.
Note that in the LDA calculation the negative exchange 
splitting has different origin: the presence of the spin-up Mn3d states
at the top of the valence band.

\begin{figure}
\caption{Total DOS and partial Mn3d-DOS for the antiferromagnetic Zn$_{1-x}$Mn$_{x}$Se
with $x=3.125\%$ calculated within LDA+U approach. 
The spin-projected DOS are calculated with respect to the local
atomic quantization axes.
\label{fig_DOS_afm_LDA_U}}
\includegraphics[width=8cm]{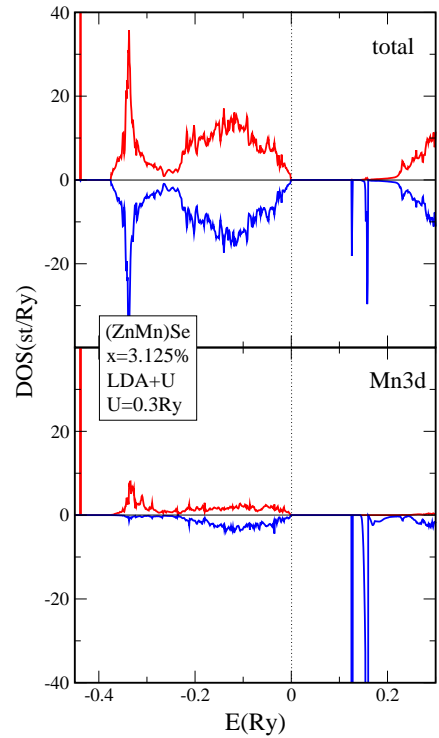}
\end{figure}

To study the relation between net magnetization and exchange splitting 
we performed calculation 
for the antiferromagnetic configuration of the Mn moments for 
$x=3.125\%$ (Fig. \ref{fig_DOS_afm_LDA_U}).
In contrast to LDA (Fig. \ref{fig_DOS_afm}), 
no exchange splitting of the states at the
top of the valence band is obtained in the antiferromagnetic case. This
property is in agreement with the mean-field picture (Fig. \ref{fig_stoner}).    
To investigate this property further we performed the LDA+U calculation 
of the exchange splitting at the $\Gamma$ point of the BZ as a function
of the Mn concentration $x$ (Fig. \ref{fig_Jpd}).
The result obtained within the LDA+U scheme is 
in very good agreement with experiment concerning both the mean-field character of
the dependence and the magnitude of the exchange parameter $J_{pd}$.  
Summarizing the study of the exchange interaction between Mn3d and valence 
band states we draw the conclusion that the LDA+U approach 
provides for this property much better agreement with experiment than the LDA approach. 

\subsection{Interatomic exchange interactions}

\begin{figure}
\caption{$T_C^{MF}$ of (ZnMn)Se 
as a function of the electron number $n$.
$n=0$ corresponds to the nominal electron number in (ZnMn)Se.
For $n=0$ there is no charge carriers in the system. 
\label{fig_Tc}}
\includegraphics[width=7.5cm,angle=-90]{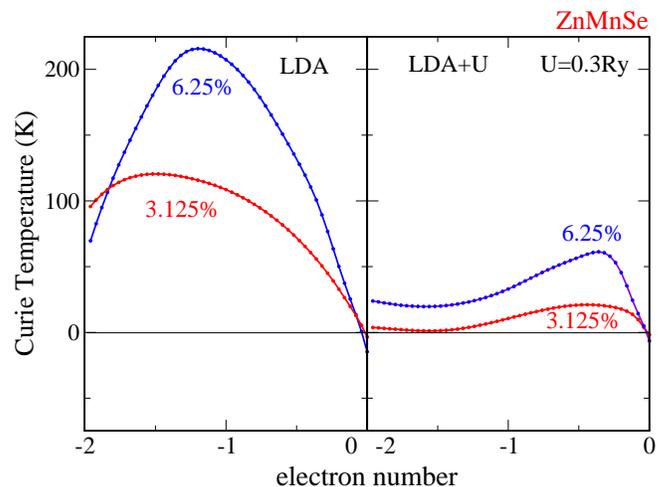}
\end{figure}

\begin{figure}
\caption{The main interatomic exchange parameters of (ZnMn)Se
with Mn concentration of $x=3.125\%$ as a function of the 
electron number. 
\label{fig_exch_par}}
\includegraphics[width=7.5cm,angle=-90]{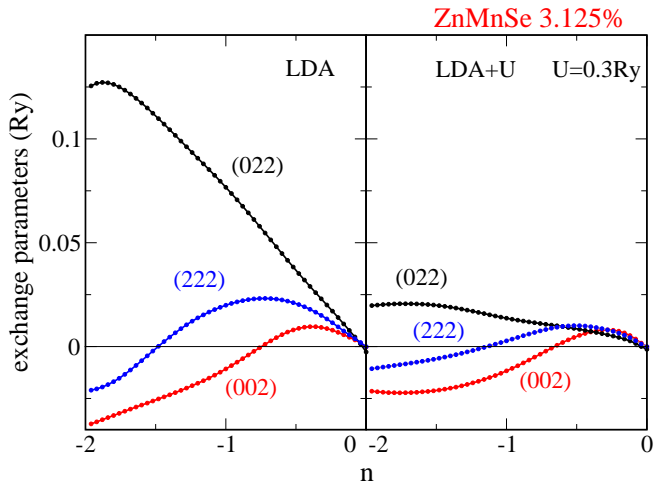}
\end{figure}

Now we turn to the discussion of the exchange interactions between Mn moments
and address the question why large Mn moments and strong $p$-$d$ exchange  
do not result in the case of (ZnMn)Se with low Mn 
content in a sizable magnetic phase-transition temperature.
To get deeper insight into formation of the interatomic exchange interactions in the system
we performed calculations for different band occupations (Fig. \ref{fig_Tc}). The number
of electrons varied from $n=-2$ (two electrons per supercell
less) to $n=0$. 
Negative values of the Curie temperature in Fig. \ref{fig_Tc}
indicate an instability of the ferromagnetic state due to dominating
antiferromagnetic interactions. The calculations have been performed within both
LDA and LDA+U approaches.
In Fig. \ref{fig_exch_par} we show, for $x=3.125\%$,
the dependence of the main interatomic exchange parameters  
as a function of the valence-band occupation.

We begin with the discussion of the features 
common for both LDA and LDA+U calculations.
Analysis of the calculated 
$T_C^{MF}$ ( Fig. \ref{fig_Tc}) shows that 
in the case of completely filled valence band and empty conduction band ($n=0$)
the main exchange interactions are antiferromagnetic. 
The antiferromagnetic character of
the interaction agrees with the commonly excepted picture that
the interatomic exchange interaction between magnetic atoms
in an insulating system is dominated by the antiferromagnetic superexchange.
This interaction is considered to be mediated by the states of the
intermediate nonmagnetic atoms \cite{magnetism_anderson} 
or the states of the completely filled bands. \cite{lahaeh}

Remarkable, however, is very small value of the exchange interactions for $n=0$.
With decreasing number of the holes the
absolute value of all interatomic exchange interactions becomes very small.
This result is in good correlation with the failure to experimentally detect
the spin-ordering up to very low temperatures. \cite{furdyna} 
It is also in agreement with the perturbative calculation by Larsen et al for 
Cd$_{1-x}$Mn$_x$Te. \cite{lahaeh}

The calculations show that the appearance of holes
results in increasing ferromagnetic interactions. This is reflected by the 
property that 
the minimum of the estimated $T_C$ is at $n=0$ (Fig. \ref{fig_Tc})
and correlates with experimental observation of the ferromagnetism 
with small Curie temperature in p-doped (ZnMn)Te. \cite{feciwa} 

Although both LDA and LDA+U give very weak negative exchange interactions for completely 
filled valence band the form of the dependence of the interatomic exchange parameters
and correspondingly Curie temperature on the number of holes differs strongly
for these two techniques
(Figs. \ref{fig_Tc},\ref{fig_exch_par}).
Comparison of the calculated exchange parameters shows
that the main difference between LDA and LDA+U concerns  
parameter $J_{022}$ that describes the exchange interaction between the Mn atoms
separated by vector (022): the LDA predicts much stronger increase of the
ferromagnetic interaction than LDA+U. The strength of the
ferromagnetic interactions obtained within LDA  seems to be
strongly overestimated since the large values
of the LDA-$T_C$ (Fig.~\ref{fig_Tc}) does not correlate with the experimental data available.
On this basis, we draw the conclusion that the LDA+U scheme gives
better description of the character of the dependence of the interatomic
exchange interactions on the number of holes.

This last conclusion 
might seem to be expected since in Sec. \ref{sec_split}
we have seen that the value of the $J_{pd}$ parameter playing an important role in the 
mediation of the ferromagnetism is overestimated by LDA.
The situation is, however, more complex since a
larger $J_{pd}$ parameter is accompanied, in the case of LDA, by a
stronger spatial localization of the valence-band hole states about the Mn atom. 
Increased localization of the 
holes produces the trend to decreasing $T_C$. Which of two factors, $J_{pd}$ or 
hole-localization, prevails depends on the peculiar interplay of the details of the electron
structure of the specific system studied. The calculations for various III-V DMS show 
that the account for Hubbard-U
can lead to both increase and decrease of the Curie temperature. \cite{sandratskii_unpubl}

To complete the consideration of the exchange interactions in (ZnMn)Se
we discuss the physical reason for a very weak interatomic exchange interactions 
for undoped (ZnMn)Se ($n=0$ in Fig. \ref{fig_Tc}). To remind, this result is common for both LDA and LDA+U 
and is very important for understanding of the failure to experimentally detect any
magnetic phase transition up to very low temperatures.
Note, that the behavior of the II-VI DMS is different from that obtained for the III-V
DMS (GaMn)As. In (GaMn)As, the same calculational scheme gives \cite{sabr_condmat} 
for the completely filled valence band and $x=3.125\%$
interatomic antiferromagnetic interactions that are
about 20 times larger that in (ZnMn)Se (Fig. \ref{fig_Tc}).

To understand the weakness of the calculated superexchange in (ZnMn)Se and
the difference between (ZnMn)Se and 
(GaMn)As we invoke a tight-binding model of noncollinear magnetic configurations. 
We consider helical configurations of the atomic
moments and study the dependence of the band energy of the system on 
magnetic structure. The helical structures are defined by the formula
\begin{equation}
\label{spir}
{\bf e}_{n} = \left( \cos ({\bf q} \cdot {\bf R}_n)
 \sin \theta,\: \sin
({\bf q} \cdot {\bf R}_n) \sin \theta,
 \:\cos \theta\right)
\end{equation}
where ${\bf R}_n$ are the lattice vectors, ${\bf q}$ is the wave vector of the helix,
${\bf e}_{n}$ is the unit vectors in the direction of the magnetic moment at site
${\bf R}_n$, polar angle $\theta$ gives the deviation of the moments from the $z$ axis.
The helical structures 
allow to describe broad range of magnetic configurations
from collinear ferromagnetism ($\theta=0$ or ${\bf q=0}$) to collinear antiferromagnetism 
(${\bf q}=\frac{1}{2}{\bf K}$ and $\theta=90^\circ$, ${\bf K}$ is a reciprocal lattice vector).

The tight binding method for spiral structures was discussed
in its general form in Ref. \cite{sand86}. 
By neglecting the difference in the
spatial dependence of the basis functions with opposite spin projections
and by preserving only the single-center matrix elements of the exchange
potential we arrive at the following simple form of the secular matrix $H({\bf k})$ \cite{sakuv}:
\begin{widetext}
\begin{equation}
\label{tb-mat}
H({\bf k})=
\left(\begin{array}{cc}\cos^2\frac{\theta}{2}\:H_-+\sin^2\frac{\theta}{2}\:H_+
-\frac{1}{2}\Delta&
-\frac{1}{2}\sin\theta\:(H_--H_+)\\
-\frac{1}{2}\sin\theta\:(H_--H_+)&
\sin^2\frac{\theta}{2}\:H_-+\cos^2\frac{\theta}{2}\:H_++\frac{1}{2}\Delta
\end{array}\right)
\end{equation}
\end{widetext}
where $H_-=H_\circ({\bf k}-\frac{1}{2}{\bf q})$, 
$H_+ =H_\circ({\bf k}+\frac{1}{2}{\bf q})$, and matrix $H_\circ({\bf k})$
describes spin-degenerate bands of a non-magnetic
crystal; $\Delta$ is the diagonal matrix of the on-site exchange splittings.
In LDA+U scheme $\Delta$ includes also $\frac{U}{2}$.
Secular matrix (\ref{tb-mat}) describes a many-band system and takes into account 
the hybridization between the states of different atoms.
Through the hybridization between the Mn states and the states
of the semiconductor matrix the spin polarization is transmitted to the 
nonmagnetic atoms. Note, that in contrast to the two-band model used 
in Ref. \cite{sabr_condmat} the matrix (\ref{tb-mat}) includes all relevant
bands. In particular the Mn3d states are assumed to be included. 
This makes model (\ref{tb-mat}) conceptually similar to our ASW calculations.

The property of the secular matrix (\ref{tb-mat}) that is important for us reads 
\begin{equation}
\label{eq:etot_full_bands}
\int_{\rm BZ} d{\bf k}Sp[H({\bf k})]=2\int_{\rm BZ} d{\bf k}H_\circ({\bf k})  
\end{equation}
that is the trace of the matrix $H({\bf k})$ does not depend on the magnetic 
configuration. 
The variation of the magnetic structure changes the energy of individual electron
states. The changes of different states, however, compensate.
In the case the trace of the secular matrix represents the total energy
of the system the invariance with respect to the magnetic configuration
means that all effective interatomic exchange interactions are negligible.

The property given by Eq. (\ref{eq:etot_full_bands}) applies 
to the total energy of a system if
all bands described by the tight-binding secular matrix (\ref{tb-mat})
are occupied. This condition can be fulfilled only in the case that 
the hybridization between the occupied and empty states is weak
and can be neglected. The weakness of the hybridization allows to
include into the secular matrix the occupied states only. 

An attempt to use  Eq. (\ref{eq:etot_full_bands}) for the explanation of 
the weakness of the superexchange in (ZnMn)Se
leads immediately to the following difficulty. According to Eq. (\ref{tb-mat}) both 
spin-up and spin-down Mn 3d states must be included into the secular matrix to fulfill 
Eq. (\ref{eq:etot_full_bands}). Since only the spin-up Mn 3d states are occupied, 
the inclusion of the
spin-down Mn 3d states, apparently, does not allow to relate the trace of the matrix to 
the energy of the system. In the case of (ZnMn)Se this difficulty can, however, be overcome
if we notice that the spin-down Mn 3d states form very narrow energy bands lying in the
semiconducting gap of ZnSe (Figs. \ref{fig_DOS},\ref{fig_DOS_U}). 
For example, the estimation for the LDA case
shows that the states of these bands are strongly localized about the Mn atoms:
more than 76\% 
is located in the Mn spheres and more than 88\% within the first coordination
sphere of the Zn atoms. These states can be treated as evanescent states that
are unable to mediate efficiently the exchange interaction between Mn atoms (Fig. \ref{fig_Tc}).
The evanescent 
character of the empty Mn 3d states allows to approximately consider their contribution
into the trace of the secular matrix [Eq. (\ref{eq:etot_full_bands})]
as being independent of the magnetic configuration. This has as a
consequence that the contribution of the occupied states is also approximately independent
of the magnetic configuration resulting in weak effective interatomic exchange
interactions.     

This consideration allows us to also explain
the difference between DMS on the 
GaAs and ZnSe bases. The semiconducting energy gap is substantially 
larger in the case of ZnSe. 
In the case of GaAs 
the influence of empty states is stronger and they must be included
into the secular matrix  (\ref{tb-mat}). Therefore, the property 
described by Eq. (\ref{eq:etot_full_bands}) does not apply to the 
occupied states that results in stronger superexchange.

Some further comments are worth making here. First, the relation between
the value of the semiconducting gap and the spatial extent of the exchange 
interactions has been many times discussed in the scientific literature 
within the framework of the perturbative treatment involving virtual 
transitions between the occupied states of the valence band and empty states of the
conduction band
(see. e.g., Ref. \cite{blro} for an early publication on this topic.
See also an interesting comment on the relation between the range of the 
exchange interaction and 
an imaginary Fermi vector in recent
paper by Pajda et al \cite{pakutu}).
The model considered above captures basically the same physics by taking into
account, in a non-perturbative manner, the hybridization between occupied
and empty states. Only in the
case the contribution of this hybridization into the response of the valence band states
on the change of magnetic configuration is small the interatomic exchange
interactions are weak. An important feature of model (\ref{tb-mat}) is that it
reflects the properties of the non-perturbative technique we used
in the calculation of the exchange interactions
and, therefore, provides additional understanding of the calculational results.

\section{Conclusions}
One of the remarkable properties of the II-VI diluted magnetic semiconductor (ZnMn)Se 
is the giant spin splitting of the valence band states under application of the 
magnetic field (giant Zeeman splitting). 
This splitting reveals strong exchange interaction between Mn moments and
semiconductor states. 
On the other hand, no magnetic phase transition has been observed 
for systems with small Mn content up to very low temperatures. The latter property shows
weakness of the exchange interaction between Mn moments.
In this paper, the local density approximation (LDA) and the LDA+U techniques are 
employed to study exchange interactions in (ZnMn)Se. 
Supercell and frozen-magnon approaches applied earlier 
to III-V diluted magnetic semiconductors 
are used. It is found that both LDA and LDA+U 
describe successfully the combination of the strong Zeeman splitting 
and weak interatomic exchange. However, the physical pictures provided by
two techniques differ strongly.
A detailed analysis shows that LDA+U method provides the description
of the system which is much closer to the experimental data.

\begin{acknowledgments}
The author profited much from the discussions with Patrick Bruno 
and Tomasz Dietl. The financial support of Bundesministerium f\"ur Bildung und
Forschung is acknowledged. 
\end{acknowledgments}

\end{document}